\documentclass[10pt]{article}
\pdfoutput=1
\usepackage{chicago,epic,graphpap,graphics,float,tabularx,paralist,longtable,fancyhdr,fancyvrb,paralist}
\usepackage{amsmath,bbm}
\usepackage{amstext}
\usepackage{amsbsy}
\usepackage{amsthm}
\usepackage{amsfonts}
\usepackage{amssymb}
\usepackage{amscd}
\usepackage{eucal}
\def\ito{It{\^o}}
\def\S{{\bf S}}
\def\m{\mu}
\def\D{\Delta}
\def\p{\pi}
\def\g{\gamma}
\def\as{\quad\text{{\rm a.s.}}}
\def\R{{\mathbb R}}
\def\eqdef{\triangleq}
\def\half{\frac{1}{2}}
\def\sumi{\sum_{i=1}^n}
\def\sumj{\sum_{j=1}^n}
\def\brac#1{\langle #1 \rangle}
\def\dd{\circ d}
\def\T{{\mathcal {T}}}
\def\Scal{{\mathcal {S}}}
\def\intT{\int_0^T}
\def\sumij{\sum_{i,j=1}^n}

\begin{document}

\centerline{\Large\bf A new decomposition of portfolio return}
\vspace{10pt} \centerline{\large  Robert Fernholz\footnote{INTECH, One Palmer Square, Princeton, NJ 08542.  bob@bobfernholz.com. The author thanks Adrian Banner, Ricardo Fernholz, Ioannis Karatzas, and Onur Ozyesil for their invaluable comments and suggestions regarding this research.}} \centerline{\today}
\vspace{10pt}

\begin{abstract}
For a  functionally generated portfolio, there is a natural decomposition of the relative  log-return  into the log-change in the generating function and a drift process. In this note, this decomposition is extended to arbitrary stock portfolios by an application of Fisk-Stratonovich integration. With the extended methodology, the generating function is represented by a {\em structural process}, and the drift process is subsumed into a {\em trading process} that measures the profit and loss to the portfolio from trading. 
\end{abstract}
\vspace{5pt}

\vspace{10pt}
\noindent{\large\bf Introduction}%%%%%%%%%%%%%%%%%%%%
\vspace{8pt}
 
For $n>1$, consider a stock market of positive, continuous, square-integrable semimartingales $X_1,\ldots,X_n$ that represent the capitalizations of the stocks. Let $\p$ be a portfolio with weight processes $\p_1,\ldots,\p_n$, which are bounded measurable processes adapted to the underlying filtration, and which add up one. Denote the total capitalization of the market  by $X(t)=X_1(t)+\cdots+X_n(t)$, and let $\m$ be the market portfolio with the market weights $\m_1,\ldots,\m_n$ such that $\m_i(t)=X_i(t)/X(t)$. Let $Z_\p$ be the value process of the portfolio $\p$, and let $Z_\m$ be the value process of the market portfolio, with $Z_\m(t)=X(t)$. More information regarding these and other definitions used in this note can be found in \citeN{F:2002} or \citeN{FK:2009}.

 A positive $C^2$ function $\S$ defined on the unit simplex $\D^n\subset\R^n$ {\em generates} a portfolio  $\p$ if 
 \begin{equation}\label{1}
d \log\big(Z_\p(t)/Z_\m(t)\big)  = d\log \S(\m(t))  + d\Theta(t),\as,
\end{equation}
where the {\em drift process} $\Theta$ is of locally bounded variation.
For an arbitrary portfolio $\p$, we shall define a {\em structural process} $\Scal_\p$, which measures the efficacy of stock selection in the portfolio, and a {\em trading process}  $\T_\p$, which measures the profit and loss from trading, such that 
\begin{equation}\label{2}
d\log\big(Z_\p(t)/Z_\m(t)\big) = d\log\Scal_\p(t) + d\T_\p(t),\as
\end{equation}
If the portfolio weights are continuous semimartingales, then $\T_\p$  will be of locally bounded variation, and for a functionally generated portfolio,
\[
d\log\Scal_\p(t)=d\log\S(\m(t))\qquad\text{ and }\qquad d\T_\p(t)=d\Theta(t),\as
\]

Let us first consider two types of stochastic integration.

\vspace{10pt}
\noindent{\large\bf \ito\ integrals and Fisk-Stratonovich integrals}%%%%%%%%%%%%%%%
\vspace{8pt}
 
We shall be considering both \ito\ integration and Fisk-Stratonovich integration, and details regarding the relationship between these two forms of stochastic integration can be found in   \citeN{protter:1990}. Let $X$ and $Y$ be continuous, square-integrable semimartingales. Then the \ito\ integral satisfies
\begin{equation}\label{3}
\intT Y(t)\,dX(t) = \lim_{\Delta\to 0} \sum_{i=1}^{\nu-1} Y(t_i)\big(X(t_{i+1})-X(t_i)\big),\as,
\end{equation}
where the limit is in quadratic mean and  $\Delta$ is the mesh of the partition $\{0=t_1<t_2<\cdots<t_\nu=T\}$.
For the {\em Fisk-Stratonovich integral,} with the differential denoted by  $\phantom{\!\!}\dd$, this becomes
\begin{equation}\label{4}
\intT Y(t)\dd X(t) = \lim_{\Delta\to 0} \sum_{i=1}^{\nu-1} \frac{Y(t_i)+Y(t_{i+1})}{2}\big(X(t_{i+1})-X(t_i)\big),
\end{equation}
where the limit is again in quadratic mean. The effect of this definition is to allow the integrand to extend past the usual filtration and be ``half-way'' into the future. It will be convenient to replace the integral in  \eqref{4} with the differential notation
\[
Y(t)\dd X(t),
\] 
as is commonly done for the \ito\ integral.

The two integrals are related by
\begin{equation}\label{5}
Y(t)\dd X(t) =  Y(t)\,dX(t) + \half \,d\brac{X,Y}_t,\as,
\end{equation}
where $\brac{X,Y}_t$ is the cross-variation process for $X$ and $Y$. Indeed, this equation is sometimes used as the definition of the Fisk-Stratonovich integral (see \citeN{protter:1990}, Chapter V). It follows from \eqref{5} that for continuous semimartingales the difference between an \ito\ integral and the corresponding Fisk-Stratonovich integral will be a process of locally bounded variation.

For a $C^2$ function $F$ defined on the range of $X$, the Fisk-Stratonovich integral satisfies the rules of standard calculus. By \ito's rule we have
\begin{equation}\label{6}
dF(X(t))=  F'(X(t))\,d X(t) + \half  F''(X(t))\,d\brac{X}_t,\as,
\end{equation}
where $\brac{X}_t$ is the quadratic variation of $X$, and since
\[
d\brac{F'(X),X}_t = F''(X(t))\,d\brac{X}_t,\as,
\]
it follows from \eqref{5} that 
\begin{equation}\label{7}
dF(X(t)) = F'(X(t))\dd X(t),\as
\end{equation}
(see \citeN{protter:1990},  Theorem~V.20).

\vspace{10pt}
\noindent{\large\bf Decomposition of portfolio return}%%%%%%%%%%%%%%%%%%
\vspace{8pt}
 
Let us consider the following thought experiment: Suppose we hold a large-capitalization stock index portfolio comprising the largest $m<n$ stocks in the market at weights proportional to their market weights. Suppose now that the stock at rank $m$ changes places with the stock at rank $m+1$, and nothing else moves.  In this case, we sell the former rank-$m$ stock and buy the current one, and after the trade the portfolio is just as it was before, except that its value has decreased. The loss in portfolio value is due to the drop in price of the original rank-$m$ stock, which was subsequently replaced by the new rank-$m$ stock. If we had been able to hold both of these two stocks, each at its average weight over the period, then the loss would have vanished.  

If we consider the portfolio log-return, the Fisk-Stratonovich integral \eqref{4} evaluates the log-return using the average weights, while the \ito\ integral \eqref{3} evaluates the log-return using the initial weights, so the difference between the values of these two integrals represents the effect of trading in our experiment.  This motivates us to use the average-weight log-return to measure the efficacy of stock selection in the portfolio, and to use the difference between the actual log-return and the average-weight log-return to measure the effect of trading.  Accordingly, we have

\vspace{10pt}
\noindent{\bf Definition 1.} For a portfolio $\p$ with value process $Z_\p$, the {\em structural process} $\Scal_\p$ is defined by
\begin{equation}\label{8}
d\log\Scal_\p(t)\eqdef \sumi \p_i(t)\dd\log \m_i(t),
\end{equation}
and the {\em trading process} $\T_\p$ is defined by
\begin{equation}\label{9}
d\T_\p(t)\eqdef d\log\big(Z_\p(t)/Z_\m(t)\big)-d\log\Scal_\p(t).
\end{equation}
\vspace{5pt}

By construction,  this definition is compatible with the return decomposition \eqref{2}.

\pagebreak
\noindent{\bf Proposition 1.} {\em If the portfolio weight processes $\p_i$ are continuous semimartingales, then the trading process $\T_\p$ will be of locally bounded variation.}
\vspace{5pt}

\noindent{\em Proof.}  From Definition~1, we have
\begin{align}
d\T_\p(t)&=d\log\big(Z_\p(t)/Z_\m(t)\big)-d\log\Scal_\p(t)\notag\\
&= \sumi\p_i(t)\,d\log\m_i(t)+\g^*_\p(t)\,dt-\sumi\p_i(t)\dd\log\m_i(t)\notag\\
&= \Big(\sumi\p_i(t)\,d\log\m_i(t)-\sumi\p_i(t)\dd\log\m_i(t)\Big)+\g^*_\p(t)\,dt,\as,\label{10}
\end{align}
where $\g^*_\p$ is the excess growth rate of $\p$ (see \citeN{F:2002}). If the portfolio weight processes $\p_i$ are continuous semimartingales, then  \eqref{5} implies that the term in the parentheses in \eqref{10} will be of locally bounded variation. Since the excess growth term is also of locally bounded variation,  so will be $\T_\p$.\qed

\vspace{10pt}
\noindent{\large\bf Decomposition of return for functionally generated portfolios}%%%%%%%%%%%%%%%
\vspace{8pt}

It was shown in \citeN{F:2002} that a positive $C^2$ function $\S$ defined on the unit simplex $\D^n$ such that for all $i$,  $x_iD_i\log\S(x)$ is  bounded on $\D^n$, will generate a portfolio  $\p$ that satisfies \eqref{1} with portfolio weights
\begin{equation}\label{12}
\p_i(t) = \Big(D_i\log\S(\m_i(t))+1 - \sumj \m_j(t)D_j\log\S(\m(t))\Big)\m_i(t),
\end{equation}
and a {\em drift process} $\Theta$  defined by
\begin{equation}\label{13}
d\Theta(t)=\frac{-1}{2\S(\m(t))}\sumij D_{ij}\S(\m(t))\,d\brac{\m_i,\m_j}_t
\end{equation}
 (see  also \citeN{KR:2015}).

\vspace{10pt}
\noindent{\bf Proposition 2.} {\em Let $\p$ be the portfolio generated by the positive $C^2$ function $\S$. Then
\[
d\log\Scal_\p(t) = d\log\S(\m(t)),\as,
\]
and
\begin{equation}\label{14}
d\T_\p(t)= d\Theta(t),\as
\end{equation}
}

\noindent{\em Proof.} Under the rules of Fisk-Stratonovich integration, 
\begin{align}
d\log\S(\m(t))&=\sumi D_i\log\S(\m(t))\dd\m_i(t)\notag\\
&=\sumi D_i\log\S(\m(t))\m_i(t)\dd\log \m_i(t)\notag\\ 
&=\sumi \p_i(t)\dd\log\m_i(t)\label{15} \\
&= d\log\Scal(t),\as,\notag
\end{align} 
by Definition~1, where \eqref{15} follows from \eqref{12} and the fact that
\[
\sumi \m_i(t)\dd\log\m_i(t)=\sumi d\m_i(t)=d\sumi \m_i(t)=0,\as
\]
With this established, 
\begin{equation*}
d\T_\p(t)= d\Theta(t),\as,
\end{equation*}
follows from \eqref{1} and Definition~1. \qed

\vspace{10pt}
\noindent{\large\bf Discussion}%%%%%%%%%%%%%%%%%%
\vspace{8pt}

We see from \eqref{12}  that the weight ratios  $\p_i(t)/\m_i(t)$  depend on the first derivatives $D_i\log\S(\m(t))$, and we see from \eqref{13} that the drift process $\Theta$ depends on the second derivatives $D_{ij}\S(\m(t))$. Hence, when changes in the market weights induce changes in the weight ratios, the effect of the weight-ratio changes will be recorded in the drift process. When a weight ratio changes, this requires  trading, so the drift process serves as a cumulative measure of the trading profit and loss. This measure is quantified by \eqref{14} of Proposition~2.  

Let us now apply Proposition~2 to calculate $\T_\p$ for some of the portfolios included in Example~3.1.6 of \citeN{F:2002}. For the market portfolio, or for any  buy-and-hold portfolio, there is no trading, and $\T_\p(t)=\Theta(t)\equiv0$. This is perhaps the minimal requirement for $\T_\p$ to be a measure of trading profit and loss --- if there is no trading, there will be no trading profit or loss. For an equal-weighted or constant-weighted portfolio, we see that $d\T_\p(t)=d\Theta(t)=\g^*_\p(t)dt$, the portfolio excess growth rate. Since the weights in these portfolios are constant, if the quadratic variation structure of the market is also constant, then $\g^*_\p$ will be constant, and so will be the rate of profit and loss from trading. Hence, at least in these simple cases, $\T_\p$ behaves in a manner consistent with expectations for a measure of trading profit and loss.

 \citeN{F:rank} introduces a class of portfolios that are generated by functions of the ranked market weights (see also  Theorem~4.2.1 of \citeN{F:2002}). Proposition~2 can be extended to these portfolios, at least with some additional regularity conditions imposed on the stock capitalization processes $X_1,\ldots,X_n$, and the interpretation of the processes $\Scal_\p$ and $\T_\p$ will remain the same as for the cases we have studied. However, for portfolios generated by functions that use more information than the values of the current  market weights $\m_i(t)$, as in \citeN{strong:2014} or \shortciteN{ssv:2016}, Proposition~2 may fail, and the processes $\Scal_\p$ and $\T_\p$  may behave in a manner that no longer corresponds to the interpretation that we have given them here.

\bibliographystyle{chicago}
\bibliography{math}

\end{document}